\documentclass[sigconf,article]{acmart}

\renewcommand\footnotetextcopyrightpermission[1]{} 
\usepackage{tikz}
\usetikzlibrary{arrows.meta,positioning}
\usepackage{tabularx}
\usepackage{listings}
\usepackage{xcolor}
\usepackage[title]{appendix}
\usepackage{graphicx}
\title{A Sequential Decision Framework for
Evaluating LLM-Guided Search}

\begin{document}

\author{Siqing Huo}
\affiliation{
    \institution{University of Waterloo}
    \country{Canada}
}
\email{siqing.huo.canada@gmail.com}
\author{Andrew Parry}
\affiliation{
    \institution{University of Glasgow}
    \country{United Kingdom}
}
\email{a.parry.1@research.gla.ac.uk}
\author{Debasis	Ganguly}
\affiliation{
    \institution{University of Glasgow}
    \country{United Kingdom}
}
\email{debforit@gmail.com}
\author{Mark Smucker}
\affiliation{
    \institution{University of Waterloo}
    \country{Canada}
}
\email{mark.smucker@uwaterloo.ca}
\author{Charles L. A. Clarke}
\affiliation{
    \institution{University of Waterloo}
    \country{Canada}
}
\email{claclark@gmail.com}

\begin{abstract}
Recent systems applying retrieval-augmented generation (RAG) and ``deep research'' iteratively generate queries, retrieve documents, and examine their contents, using the resulting information to guide subsequent search steps.
When this process is carried out autonomously rather than by a human searcher, the system must make explicit decisions about which search trajectories to explore and what objective the search is intended to optimize.
We introduce a simple framework for reasoning about such search policies, providing a principled approach to evaluation.
The framework models the search environment as a set of ranked lists and represents behavior using two primitive actions: \emph{select}, which advances judgment through an existing ranking, and \emph{spawn}, which generates a new ranking through query generation or reformulation.
Each action produces a document to inspect, yields information that updates the search state, and contributes a reward for a task-specific objective.
Within this abstraction, a variety of existing retrieval strategies can be expressed as policies operating over a common environment, while common retrieval objectives can be represented through reward and weighting functions over the resulting search trajectory, and the resulting formulation admits a natural reinforcement learning interpretation.
We illustrate the approach using high-recall information retrieval (HRIR), where the goal is to identify as many relevant documents as possible with limited judging effort, showing how the framework provides a unified way to describe and reason about strategies for both classical retrieval tasks and emerging LLM-guided search systems. 
Through experiments, we demonstrate that both the underlying retrieval method and the "deep research" policy employed can materially affect performance.
\end{abstract}

\pagestyle{plain}

\maketitle

\section{Introduction}

Recent systems for retrieval-augmented generation (RAG) and
``deep research'' exhibit a common structure: They iteratively generate
queries, retrieve documents, examine their contents, and use the
resulting information to guide subsequent search steps~\cite{lewis2020rag,izacard2021fid,pradeep2024ragnarok,gan2025rageval,yao2023react,meng2026revisiting}. 
This process
resembles what has sometimes been called ``slow search,'' in which a
system incrementally explores multiple search trajectories rather than
relying on a single static ranking \cite{teevan2013slow}.  From this
perspective, modern research agents can be viewed as policies operating
in a search environment: They generate new rankings, decide which
documents to inspect, and accumulate evidence over time.

This pattern is closely related to earlier work on
high-recall information retrieval (HRIR), which addresses the problem
of identifying as many relevant documents as possible within a large
collection~\cite{cormack1998efficient,pickens2010active,cormack2015autonomy,cormack2016totalrecall,losada2017bandits,hosseini2012uncertainty}.  Historically, HRIR has relied on human assessors to judge
documents and guide the search process, which makes it closely related to
interactive retrieval.  In such settings, the system produces ranked
lists, while the human decides which documents to examine next and
when to reformulate the query.  Recent advances in large language
models (LLMs) change this picture by enabling automated judgments and
query generation, allowing the search process to be controlled by an
algorithmic policy rather than a human user.  Viewed in this way,
search becomes the problem of repeatedly deciding where to allocate
effort in order to discover useful information as efficiently as
possible.

The framework described in this paper provides a simple abstraction for reasoning about such policies and a principled basis for evaluating them.
We view the search process as a
sequential decision problem in which the system repeatedly chooses how
to allocate the judging effort.  At any point, the system is in a search
state summarizing the current environment, including the ranked lists
that have been generated, the documents already examined, and any
information obtained from previous judgments.  A policy maps this
state to an action that determines which ranking should be explored
next.

We represent these decisions using two primitive actions.  The
\textbf{select} action chooses an existing ranked list and advances
judgment within that list, examining the next unseen document.  The
\textbf{spawn} action generates a new ranked list, for example, by
issuing a new query, applying a different ranking function, or
performing a query reformulation.  Each action ultimately identifies
the next document to be judged, producing an observation that updates
the search state and may yield a reward signal reflecting the value
of the information discovered. By expressing the search process in
terms of these two actions, a wide range of existing retrieval methods
can be described as policies operating over a common environment, and
the same abstraction also enables the systematic design and evaluation
of new policies for LLM-guided search. 

After a review of related work in Section~\ref{sec:related},
Section~\ref{sec:framework} presents the proposed framework and
formalizes the search process as a sequential decision problem over
ranked lists.  Section~\ref{sec:rewards} and Section~\ref{sec:policies} show how a variety of existing retrieval strategies can be expressed as policies within this framework.
For the examples and experiments in this paper, we focus on
high-recall information retrieval (HRIR), using relevance-based reward
functions and evaluating policies in terms of recall-oriented metrics.
Section~\ref{sec:experiments} presents empirical results comparing a
range of policies under this setting. Finally, we position our framework as a first step towards a general "learning to search" framework for LLM-guided search.

\begin{table*}[t]
\centering

\renewcommand{\arraystretch}{1.15}
\begin{tabularx}{\textwidth}{l|l|X|X}
\hline
Policy & Section & \textit{spawn}: Generate ranking and judge & \textit{select}: Judge from an existing ranking \\
\hline\hline

adhoc
& \S\ref{sec:adhoc}
& Executed once to generate the initial ranked list
& Always execute $\textit{select}(1)$, observing documents sequentially in rank order
\\
\hline

MTF
& \S\ref{sec:MTF}
& Draw a new ranked list from an external fixed pool when the end of the active queue is reached and additional lists remain available
& Always select the current list in queue order; if the resulting observation is relevant, move that list to the front of the queue
\\
\hline

Thompson
& \S\ref{sec:Thompson}
& When Thompson sampling selects an arm that has not yet been introduced into the environment, execute $\textit{spawn}$ to create the corresponding list and observe its first document
& Otherwise, execute $\textit{select}(i)$ for the active list whose sampled posterior value is largest
\\
\hline

ISJ
& \S\ref{sec:ISJ}
& Generate a new query from the current search context (historically by a human assessor; in our experiments potentially by an LLM) and execute it to create a new ranked list
& Select from an existing active list to continue observing documents from previously generated rankings
\\
\hline

CAL
& \S\ref{sec:CAL}
& Retrain a classifier on the labeled set accumulated so far, and apply it to the collection to produce a new ranked list whenever a new model becomes available
& Continue selecting from the most recently available ranking until a newer ranking has been spawned, then switch to that newest list
\\

\end{tabularx}
\caption{Summary of policies considered in this paper, expressed in
terms of the $\textit{spawn}$ and $\textit{select}$ actions introduced in
Section~\ref{sec:framework}. Classical ad hoc retrieval corresponds
to a single initial ranking that is always selected; MTF and Thompson
sampling allocate effort across multiple rankings; ISJ introduces new
rankings through query reformulation; and CAL introduces new rankings
through repeated retraining and reranking.}
\label{tab:policies}
\end{table*}

\section{Related Work}
\label{sec:related}

Our framework builds upon and connects three lines of related work: high-recall retrieval, interactive and exploratory search, and emerging LLM-guided search systems.

\subsection{High-Recall Information Retrieval}

Previously, high-recall retrieval has been studied extensively in the context of technology-assisted review and the ``total recall'' problem~\cite{cormack2016totalrecall}.
Work on active learning also provides important methodological foundations, from early uncertainty sampling for text classification~\cite{lewis1994sequential} to explicit active-learning approaches for high-recall retrieval~\cite{pickens2010active}.
Subsequent work on continuous active learning emphasized reliable end-to-end review, including recall-oriented variants and scalability analyses for technology-assisted review~\cite{cormack2015multifaceted,cormack2016scalability,cormack2015autonomy}.
Other related work examined the reliability of assessments and the practical conduct of total-recall review~\cite{cormack2017roger,mustafa},
as well as interaction choices such as judging short excerpts rather than full documents~\cite{zhang2018effective}.

Evaluation in high-recall retrieval is naturally defined relative to a judging budget $J$, where $J$ denotes the number of documents that may be assessed during the search process.
In practice, $J$ may be set operationally, for example, by available time, assessor effort, or a fixed review budget.
However, for evaluation, a useful normalized approach is to define the judging budget as a function of the number of known relevant documents for the topic, for example $J=aR+b$, where $R$ is the number of known relevant documents~\cite{grossman2016trec}.
Although this approach does not represent viable stopping rules in practice, since it depends on advanced knowledge of $R$, it avoids distortions that arise from applying the same fixed judging budget to topics with very different numbers of relevant documents.
The special case $J=R$ corresponds to recall@$R$, or equivalently R-precision. 
In Section~\ref{sec:experiments}, we adopt this normalized approach and report results using the special case $J=R$.

\subsection{Interactive and Exploratory Search}

More broadly, the iterative exploration of search results and the reformulation of queries are closely related to the work on interactive and exploratory search.
Those traditions emphasize that search is often open-ended, iterative, and opportunistic rather than a simple query-response transaction~\cite{pirolli2007information,belkin1998trec6,belkin1999trec8,marchionini2006exploratory,white2009exploratory,kelly2009interactive}.
This perspective is also reflected in the work on ``slow search,'' which explicitly treats search as a process that unfolds over time rather than one constrained to an immediate response~\cite{teevan2013slow}.
In such settings, users examine the results, follow promising directions, and reformulate their queries as their understanding develops.
Related work has also modeled search as a process in which users decide when to continue along a current direction and when to switch to a new one~\cite{smucker2016switching}.
Taken together, this line of work views search as a process unfolding over time across multiple possible trajectories rather than as the passive consumption of a single static ranking.
Taking this human search process as a starting point, we can recast it in the language of sequential decision making and describe search as a policy-driven process.

\subsection{LLM-Guided Search}

Recent work on retrieval-augmented generation (RAG) has shown how
retrieved documents can support generation, initially with relatively simple pipelines in which one or a small number of queries are used to retrieve supporting material for downstream synthesis~\cite{lewis2020rag,izacard2021fid,pradeep2024ragnarok}.
Retrieval may also follow generation, for example, to retrieve supporting evidence or validate factual claims in a response~\cite{huo2023retrieving,gao2022posthoc}.
More recent systems extend this pattern into multi-step search processes that iteratively generate queries, retrieve documents, and use accumulated evidence to guide subsequent search steps, moving toward what are now often described as ``deep research'' agents~\cite{futuresearch2025,agentIR,xi2025surveyllmbaseddeepsearch,hu2026sage,shi2026eigentsearchqplus,li-etal-2025-search-o1}.
Some of this recent work has begun to optimize deep search agents directly using reinforcement learning and intermediate process rewards~\cite{jin-etal-2025-search-r1,zheng-etal-2025-deepresearcher,subsearch2026}, reinforcing the need for a framework in which such search policies can be described and compared.

Although our experiments in this paper focus on high-recall retrieval, the same general framework could also be applied to settings in which the objective is not to find all relevant documents, but to discover informational ``nuggets'' or aspects of an information need~\cite{voorhees2005qa,lin2007deconstructing,clarke2008novelty,santos2010xquad,yilmaz2010ebu,sakai2011intent,lajewska2025ginger,pradeep2024ragnarok,pradeep2025greatnugget}.
This perspective also connects to earlier work on adaptive allocation of judging effort, including bandit-based approaches to test collection construction and document assessment~\cite{bandits19}, suggesting a broader continuity between classical adaptive retrieval and emerging LLM-guided search systems.

To support the evaluation of such systems, new benchmarks have been proposed that measure their ability to answer complex questions after conducting multi-step search. For example, BrowseComp-Plus~\cite{chen2025browsecompplus,sharifymoghaddam2026projectingbrowsecompplusclimbmixrealistic} provides a controlled benchmark for evaluating deep research agents, focusing on end-to-end task performance, including final answer accuracy. While such benchmarks assess the effectiveness of the overall agent, they do not explicitly model the decisions that control the search process. The framework proposed here instead focuses on those decisions, enabling analysis of the policies that govern how search trajectories are explored.

\section{Framework}
\label{sec:framework}

We begin by considering a query $Q$ that initiates a search process over a
large document collection.  As discussed in the introduction, such a
process may arise in a variety of settings, including classical
high-recall retrieval, retrieval-augmented generation (RAG), and
LLM-guided ``deep research.''  In each case the system repeatedly
retrieves documents, examines their contents, and uses the resulting
information to guide subsequent search steps.  Our goal is to discover
useful information as efficiently as possible while exploring the
space of possible queries and rankings.  The value of the information
obtained at each step is captured by a reward signal derived from the
documents examined.  Depending on the task, this reward may reflect
binary relevance, graded gain, or the discovery of informational ``nuggets''.  

We model this search process as the incremental exploration of a growing set of ranked lists.
At each time step $t$ the system maintains a set of ranked lists, with initially $L = \{\}$ at $t = 0$:
\[L = \{L^1, L^2, \dots, L^k\}.\]
Each list $L^i$ is an ordered sequence of documents
\[L^i = (d^i_1, d^i_2, d^i_3, \dots)\]
where $d^i_j$ denotes the document
at rank $j$ in list $i$.
Each list maintains a depth pointer $p^i$ indicating that documents
$d^i_1 \dots d^i_{p^i}$ have already been ``observed'' or ``judged''.
The next candidate document from list $i$ is therefore $d^i_{p^i+1}$.

Let $E_t$ be the set of documents judged after step $t$.
When document $d$ is judged, the environment produces an observation
$\mathcal{O}(d)$ describing the information obtained from
that document.  Depending on the task, this observation may represent
binary relevance, graded gain, or the discovery of informational
nuggets. 
A reward function $r(\mathcal{O}(d))$ maps this observation
to a scalar reward, reflecting the discovery of a new relevant document or a new nugget.
At step $t$, the environment contains the current lists $L$,
their depth pointers $p^1 \dots p^k$, and any additional information
such as scores, ranks, or previously observed judgments.
We denote the complete environment state abstractly by $\theta_t$.

At each step $t$, the system performs one of two actions:
selecting the next document from an existing list, or spawning a new ranked list:
\begin{description}

\item[\textbf{Select}]
Given the current environment state $\theta_t$, select a list index $i$
and observe the next unseen document from list $i$.
Starting at position $p^i+1$, previously observed documents are skipped
until an unseen document $d^i_j$ is found.
The system observes $\mathcal{O}(d^i_j)$, updates the pointer
$p^i \leftarrow j$, and adds $d^i_j$ to $E_t$.
Depending on the task, $\mathcal{O}(d^i_j)$ may correspond to a
binary or graded relevance judgment, or the discovery of informational
nuggets, and may produce a reward $r(\mathcal{O}(d^i_j))$.

\item[\textbf{Spawn}]
Create a new ranked list by generating $L^{k+1}$ and initializing
$p^{k+1}=0$.
A \textit{spawn} operation may correspond to generating a new query,
applying a different ranking function, or another retrieval strategy.
After creating the new list, the system immediately executes
$\textit{select}(k+1)$.

\end{description}
The actions $\textit{select}(i)$ and $\textit{spawn}$ correspond,
respectively, to exploiting existing rankings and exploring new
search trajectories.
Each step produces exactly one observation, so that $t$ corresponds
to the number of observations made\footnote{
In our model, an observation $\mathcal{O}(d)$ corresponds to a single LLM judgment event, which we treat as a unit cost analogous to one human judgment in a traditional search process.
This abstraction simplifies the true costs in several ways.
Modern retrieval pipelines may also use LLM calls before a document is observed, for example, for pre-filtering or reranking, including approaches such as pairwise reranking and listwise reranking~\cite{nouriinanloo2024reranking,qin2024pairwise,zhuang2024setwise,sun2023rankgpt,pradeep2023rankvicuna,pradeep2023rankzephyr}.
It also ignores cost variation across prompts due to differing input and output token counts.
Finally, at this level of abstraction, we treat retrieval and ranking costs as negligible relative to LLM judgment costs.
}

Both operations are controlled by a policy $\pi(\theta_t)$
which maps the current environment state to either
$\textit{spawn}$ or $\textit{select}(i)$.
The goal of the policy is to maximize the cumulative reward obtained
during the search process, which may be expressed as a discounted sum:
\begin{equation}
\sum_{t=1}^{\infty} \gamma^{t-1} r_t ,
\end{equation}
where $r_t$ is the reward obtained from the observation at step $t$
and $\gamma$ is a discount factor.
For, search, it is more convenient to use a general
weighting of rewards over time:
\begin{equation}
\label{eqn:reward}
   \sum_{t=1}^{\infty} w(t)\, r_t , 
\end{equation}
where $w(t)$ is a weighting function that reflects the value of
discovering useful information earlier in the search process~\cite{wang2019beyond}.
As we will see in the next section, many common IR evaluation measures can be expressed in this form~\cite{jarvelin2002cumulated,clarke2008novelty}.

\section{Rewards}
\label{sec:rewards}

Equation~\ref{eqn:reward} provides a general form for expressing many information retrieval evaluation measures, generalizing them to our framework.
To illustrate this, consider a simple policy in which the system
executes $\textit{spawn}$ once to create a ranked list and then
repeatedly executes $\textit{select}(1)$, observing documents
sequentially in rank order.
In this case the trajectory is determined entirely by the order of the ranking, and different evaluation measures arise from different choices of the weighting function $w(t)$ and reward $r_t$.

For example, discounted cumulative gain (DCG) corresponds to choosing a
logarithmic discount
\begin{equation}
    w(t) = \frac{1}{\log_2(t+1)}
\end{equation}
with $r_t$ defined as a gain value derived from the document's
relevance label, often using an exponential gain such as
$r_t = 2^{\mathrm{rel}_t} - 1$.  Rank-biased precision (RBP) may be
expressed using a geometric discount
\begin{equation}
  w(t) = (1-\rho)\rho^{t-1},
\end{equation}
where $\rho$ is a persistence parameter, and $r_t$ typically represents
binary relevance in the range $[0,1]$.  Many other ranking measures can
be written in the same form by selecting appropriate gain and weighting
functions.

In this paper we focus primarily on high-recall information retrieval
(HRIR), where the goal is to discover as many relevant documents as
possible while minimizing judging effort.  In this setting, the weighting
function reflects a judging budget.  If the available judging budget is
$J$, we define
\begin{equation}
w(t) =
\begin{cases}
1 & \text{if } t \le J, \\
0 & \text{otherwise.}
\end{cases}
\label{eqn:budget}
\end{equation}
In this case, the reward $r_t$ is simply a binary indicator of relevance
for the document observed at step $t$.  The resulting objective
corresponds to maximizing the number of relevant documents discovered
within the judging budget.

The same formulation can be extended to alternative reward definitions.
For example, in tasks that emphasize information coverage rather than
document relevance, rewards may correspond to the discovery of
informational nuggets.  In this case $r_t$ indicates whether the
observation reveals previously unseen information, and the objective in
Equation~\ref{eqn:reward} measures the cumulative information gained
during the search process.

\section{Policies}
\label{sec:policies}

In this section, we map existing work into the framework of
Section~\ref{sec:framework}, with a continued focus on HRIR.
The policies are summarized in Table~\ref{tab:policies}.

\subsection{Adhoc Search}
\label{sec:adhoc}

The simplest policy corresponds to the classic adhoc search scenario.
The system first executes $\textit{spawn}$ to generate a ranked list
and then always selects that list.
At each step the system executes $\textit{select}(1)$, observing
$d^1_1, d^1_2, d^1_3, \dots$ in rank order.
The process continues until $J$ observations have been made.
In the HRIR setting, where $r_t$ is a binary indicator of relevance
and the weighting function enforces a judging budget of $J$,
the cumulative reward corresponds to \emph{Recall@J}.

Classic adhoc search provides natural baselines for the
experiments in Section~\ref{sec:experiments}.  We consider multiple
rankers to generate the ranked lists, including BM25, sparse learned,
and dense retrieval methods, as well as a hybrid two-stage retrieval pipelinem where the results from all three methods are pooled and reranked with MonoT5 and DuoT5~\cite{monoduo}.

\subsection{Move-to-Front}
\label{sec:MTF}

The Move-to-Front (MTF) strategy~\cite{cormack1998efficient} assumes a
fixed pool of ranked lists produced by different retrieval systems.
These lists exist outside the search process and are introduced into
the environment through the $\textit{spawn}$ action.
The environment maintains the active lists in an ordered queue.
Initially $L=\{\}$, so the first action must be $\textit{spawn}$,
which introduces a ranked list and observes its first document.
At each step the policy executes $\textit{select}(i)$ where $i$
corresponds to the current position in the queue.  The next unseen
document from that list is observed.  If the observation indicates a
relevant document, list $i$ is moved to the front of the queue.
Otherwise the next step considers the following list in the queue.
If the end of the queue is reached without observing a relevant
document, the policy executes $\textit{spawn}$ to introduce a new
ranked list from the external pool if one is available; otherwise the
scan resumes from the beginning of the queue.
Intuitively, lists that produce relevant documents are promoted to the
front so that they will be exploited on subsequent steps, while new
lists are introduced only when the existing rankings fail to yield
additional relevant material.

\subsection{Thompson Sampling}
\label{sec:Thompson}

The Move-to-Front policy may be viewed as a simple heuristic solution
to the exploration--exploitation problem.
Each ranked list acts as an
``arm'' whose value is revealed through observations obtained by
executing $\textit{select}(i)$.  Lists that recently produce relevant
documents tend to remain near the front of the queue and are therefore
selected more frequently.  From this perspective, MTF behaves like a
lightweight multi-arm bandit policy for allocating judging effort
across multiple rankings.
In addition to MTF, other bandit-style approaches have previously been explored in the context of HRIR for adaptive test collection construction \cite{bandits19,losada2017bandits}.

Once the problem is viewed though the lens of multi-arm bandits, it is natural to consider standard bandit algorithms.  Thompson sampling~\cite{thompson1933likelihood}
provides a simple and effective alternative.  Each ranked list is
associated with a probability distribution representing the likelihood
that the next unseen document from that list will be relevant.  At each
step a value is sampled from the distribution associated with each arm,
and the policy selects the arm with the highest sampled value.

In our framework the arms correspond to ranked lists drawn from an
external pool.  If the selected arm has not yet been introduced into
the environment, the policy executes $\textit{spawn}$ to create the
corresponding list and observe its first document.  Otherwise the
policy executes $\textit{select}(i)$ to observe the next unseen
document from the list.  In either case the resulting observation
updates the posterior distribution associated with that arm.
This formulation preserves the accounting used throughout the paper,
where each action produces exactly one observation, while providing a
principled mechanism for balancing exploration and exploitation across
multiple rankings.

\subsection{Interactive Search and Judging}
\label{sec:ISJ}

Interactive Search and Judging (ISJ)~\cite{cormack1998efficient} was
originally proposed as a human-in-the-loop search process designed to
support high-recall retrieval tasks.  In the original formulation, a
human assessor examines documents returned by a query, identifies
relevant material, and periodically reformulates the query to discover
additional relevant documents.  The system therefore alternates between
judging documents from existing rankings and generating new queries
that may lead to additional relevant material.

In the framework of Section~\ref{sec:framework}, this behavior can be
expressed using the $\textit{select}$ and $\textit{spawn}$ actions.
The ranked lists correspond to the results of queries issued during the
search process.  Executing $\textit{select}(i)$ observes the next unseen
document from an existing ranking, while $\textit{spawn}$ generates a
new query and introduces the resulting ranked list into the environment.

Historically the equivalent of the $\textit{spawn}$ action was performed by a human
assessor who reformulated queries based on the evolving search context.
In the present framework, however, this role may be performed by an
automated policy.  For example, a new query may be generated by
prompting a language model with information from the environment
$\theta_t$, such as previous queries, examples of relevant documents,
and examples of non-relevant documents.  The resulting query is then
submitted to the retrieval system, producing a new ranked list that
can be explored by subsequent $\textit{select}$ actions.  Section~\ref{sec:experiments}
demonstrates one such approach using LLM-generated queries.

\subsection{Continuous Active Learning}
\label{sec:CAL}

Continuous Active Learning (CAL)~\cite{cormack2015autonomy,mustafa} was
originally proposed as an automated strategy for technology-assisted
review in high-recall retrieval tasks.  In CAL, documents are judged
sequentially while a classifier is repeatedly retrained on the growing
set of labeled documents.  After each update, the classifier is applied
to the entire collection to produce a new ranking from which subsequent
documents are selected.

In the framework of Section~\ref{sec:framework}, training the classifier
and applying it to the collection corresponds to executing
$\textit{spawn}$, which generates a new ranked list based on the
current environment state $\theta_t$.  Because training may take
nontrivial time, the system continues to observe documents from the
current ranking until a new ranking becomes available.

The simplest CAL policy always executes $\textit{select}(i)$ where $i$
corresponds to the most recently spawned list.  When a new model
finishes training, the resulting ranking is introduced by executing
$\textit{spawn}$, after which the policy begins selecting documents
from the new list~\footnote{
We may think of this policy as spawning as soon as possible,  while continuing to judge documents from the most recent available list in the meantime.
The frequency of spawning depends on the relative time to retrain a model compared to the time to judge a document.
With a human judge, this policy works in practice, but with an LLM-based judge the relative timing may differ substantially, potentially changing the effective spawning rate.}.
Thus the search process alternates between judging
documents from the current ranking and spawning new rankings produced
by retraining the classifier on the accumulated judgments.
In this interpretation the $\textit{spawn}$ operation is triggered by
the learning procedure, and the availability of a new list, rather than by an explicit policy decision,
while the policy itself simply selects documents from the most recent
ranking.

\subsection{Query Reformulation Policies}
\label{sec:spawn}

In many search processes the key step that advances the search is
query reformulation.  A human searcher typically gathers information
from a set of documents and then reformulates the query to pursue a
new direction.  One query leads to another as the searcher gradually
discovers new aspects of the information need.  In high-recall search,
this process is particularly important: relevant documents often
appear in different regions of the collection that can only be
reached through new queries. In deep research, finding new informational nuggets largely requires new queries.

In the framework of Section~\ref{sec:framework}, this behavior is
captured by the $\textit{spawn}$ action.  Executing $\textit{spawn}$
generates a new ranked list based on the current environment state
$\theta_t$.  In practice this often corresponds to generating a new
query derived from the current search context and submitting it to a
retrieval system.  The resulting ranking introduces a new search
trajectory that may reveal additional relevant documents.

Recent work has explored automated approaches to query reformulation,
particularly using large language models (LLMs).  For example, QueryGym~\cite{bigdeli2025querygym} provides an extensible framework for implementing and evaluating query reformulation
methods. 
The toolkit supports a variety of LLM-based reformulation
strategies, integrates with common retrieval backends such as
Pyserini~\cite{lin2021pyserini} and PyTerrier~\cite{macdonald2020pyterrier},
and provides infrastructure for reproducible experimentation.
As an alternative to dedicated query reformulation approaches, prompt optimization frameworks can be used to generate reformulations on the fly from the current search context, including previous queries and examples of relevant and non-relevant information.

Within the framework described in this paper, query reformulation methods can be viewed as implementations of the $\textit{spawn}$ operator.
Given the current environment state $\theta_t$, a reformulation method may generate a new query that is submitted to the retrieval system to produce a new ranked list. 

Importantly, query reformulation specifies how new queries may be generated, but not when they should be generated. 
The decision to execute $\textit{spawn}$ remains part of the policy $\pi(\theta_t)$ controlling the search process.
This separation allows the framework to combine different reformulation methods with different policies, including
policies that trigger reformulation periodically, policies that react to observed relevance signals, and policies that learn when spawning new queries is most beneficial.

\subsection{Policy Design Space}
\label{sec:extensions}

The framework above exposes a natural design space for search
policies.  Once the search process is expressed in terms of the
$\textit{spawn}$ and $\textit{select}$ actions, a variety of strategies
can be constructed by borrowing ideas from reinforcement learning,
bandit algorithms, and simple search heuristics~\cite{auer2002finite,li2010contextual}.  The following
examples illustrate several possibilities.

\begin{itemize}

\item \textbf{Upper Confidence Bound (UCB).}
Each ranked list can be treated as an arm in a multi-armed bandit~\cite{auer2002finite}.
A UCB-style policy selects the list with the largest optimistic
estimate of relevance probability, balancing lists that have produced
relevant documents against lists that remain uncertain.

\item \textbf{$\varepsilon$-greedy selection.}
A simple baseline selects the list with the highest observed relevance
rate most of the time, but occasionally selects a random list with
probability $\varepsilon$, ensuring continued exploration.

\item \textbf{Spawn triggers.}
Policies may introduce explicit conditions under which new rankings
are spawned.  For example, a new query may be generated when no
relevant documents have been observed for several judgments, when the
estimated relevance probability of all active lists falls below a
threshold, or at fixed intervals during the search process.

\item \textbf{Contextual bandits.}
Rather than treating each ranking as a single arm, the policy may use features of the environment state $\theta_t$ such as document scores, rank positions, or previous judgments to estimate the probability of relevance.
Contextual bandit algorithms can then guide selection~\cite{li2010contextual}.

\item \textbf{Depth-aware policies.}
Relevance probability typically decreases with rank depth.  Policies
can incorporate this information by estimating relevance probability
as a function of both the list and the current depth pointer $p^i$,
allowing the system to move more quickly away from exhausted rankings.

\item \textbf{Hybrid policies.}
Policies may combine bandit-based selection with spawning mechanisms
that generate new rankings through query reformulation, alternative
ranking functions, or classifier retraining.

\end{itemize}

These examples illustrate that once the search process is expressed in
terms of $\textit{spawn}$ and $\textit{select}$ actions, many existing
techniques can be adapted to produce new policies for high-recall
retrieval.

\section{Experiments}
\label{sec:experiments}

Our experiments illustrate how selected policies from Table~\ref{tab:policies} can be instantiated within the framework,
demonstrating how the framework can be instantiated with real systems and how different policies can be implemented and compared in practice.
The experiments are based on the TREC Deep Learning 2019 collection.
We use GPT-4o-mini as the LLM for relevance judgments and GPT-5 for more complex tasks such as spawning.
In our experiments, spawning is implemented through query reformulation.
We employ DSPy \cite{khattab2024dspy} for query reformulation to avoid the need to create manual prompt boilerplate, while enabling the background optimization of instructions and few-shot examples.

Each relevance judgment corresponds to a step in Equation~\ref{eqn:reward}.
Following the approach of \citet{grossman2016trec} to compare high-recall  retrieval methods, we define the judging budget as a function of the number of known relevant documents $R$ for each query.\
This approach provides normalization, so that recall values can be meaningfully averaged across queries with a wide range of relevant documents.
After a search is completed, the average recall@R is reported using official human relevance judgments to evaluate performance. 

\begin{itemize}
    \item \textbf{Adhoc Search} (\S\ref{sec:adhoc}): This is our baseline policy where we only execute $\textit{select}$ on a single ranked list. We keep judging from the top of the ranked list until we run out of judging budget.
    \item \textbf{Query Reformulation} (\S\ref{sec:spawn}) \textbf{then MTF} (\S\ref{sec:MTF}): We have a fixed pool of multiple ranked lists and only execute $\textit{select}$ until we run out of judging budget. In this experiment, the initial pool of ranked lists is generated using batch query reformulation, where we generate a list of distinct queries all at once given the original query and its top retrieved documents (Appendix~\ref{app:batch_query_reform}). Then we decide $i$ of $\textit{select}(i)$ using the MTF strategy.
    \item \textbf{Forward Only Spawning}: Always executes $\textit{select}(i)$ where $i$ corresponds to the most recently spawned list. After a new ranking list is introduced by executing $\textit{spawn}$, the policy begins only to select documents from the new list. Once we decide a ranking list is exhausted (in this experiment, we define it to be 5 irrelevant documents found), we $\textit{spawn}$ by generating a new query using the cumulative context of past queries and their top retrieved results (Appendix~\ref{app:divergent_sig}).
    \item \textbf{Bounded Spawning}: Because $\textit{spawn}$ can be expensive, we enforce a fixed quota on the number of $\textit{spawn}$ actions. Once the limit is reached, $\textit{spawn}$ is no longer allowed and $\textit{select}$ is no longer limited to only the most recently spawned ranking list. We use $\textit{select}$ with the MTF strategy to merge what is left of each ranking list.
\end{itemize}

For each type of policy, we experiment with four retrieval pipelines: BM25, SPLADE-v3 \cite{splade}, TCT-ColBERT-v2 \cite{lin-etal-2021-batch}, and a hybrid two-stage system that pools the results of BM25, SPLADE and TCT-ColBERT and reranks the pooled results with MonoT5 and DuoT5~\cite{monoduo}.

\begin{figure*}[t]
    \centering
    \includegraphics[width=0.8\textwidth]{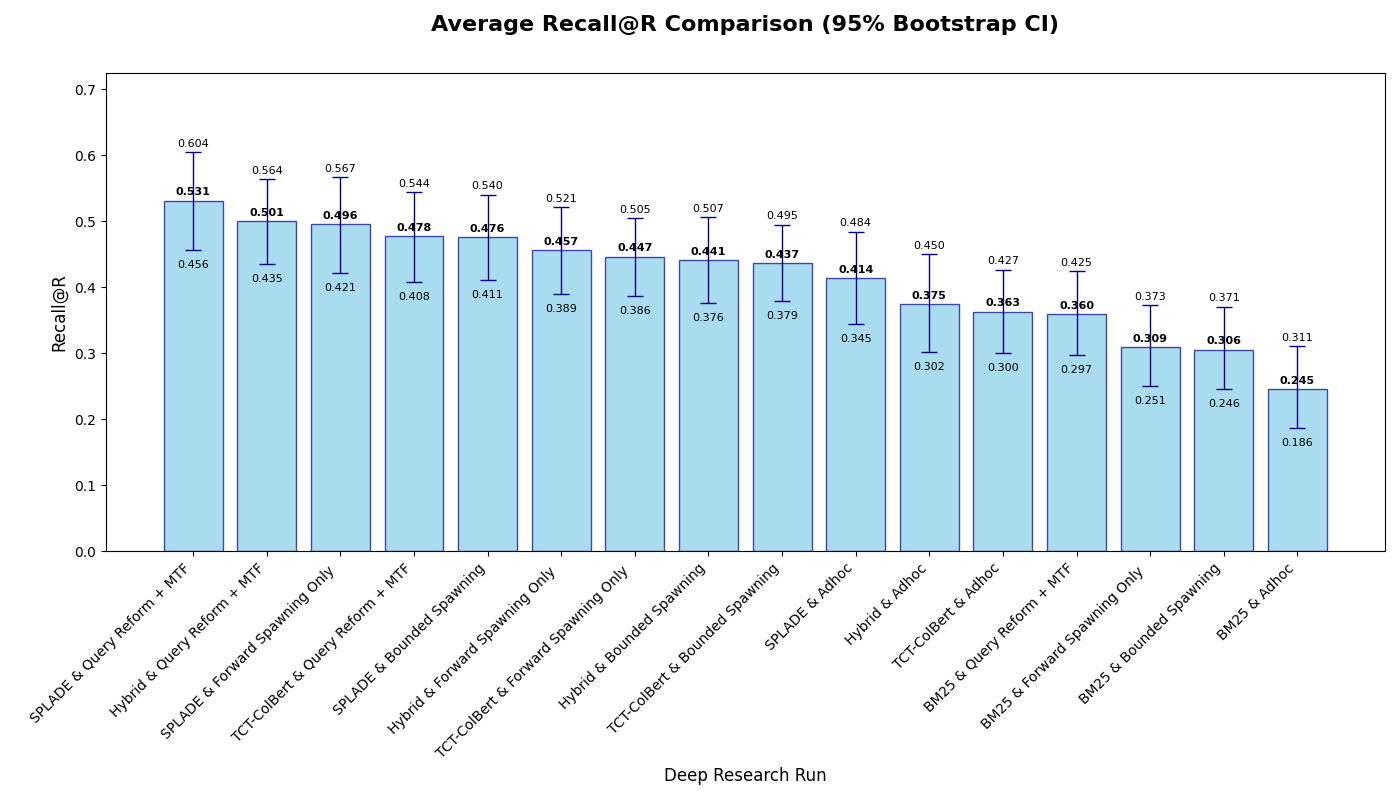} 
    \caption{Comparison of recall@$R$ across different search runs.}
    \label{fig:recall_comparison}
\end{figure*}

\begin{figure}[t]
    \centering
    \includegraphics[width=0.4\textwidth]{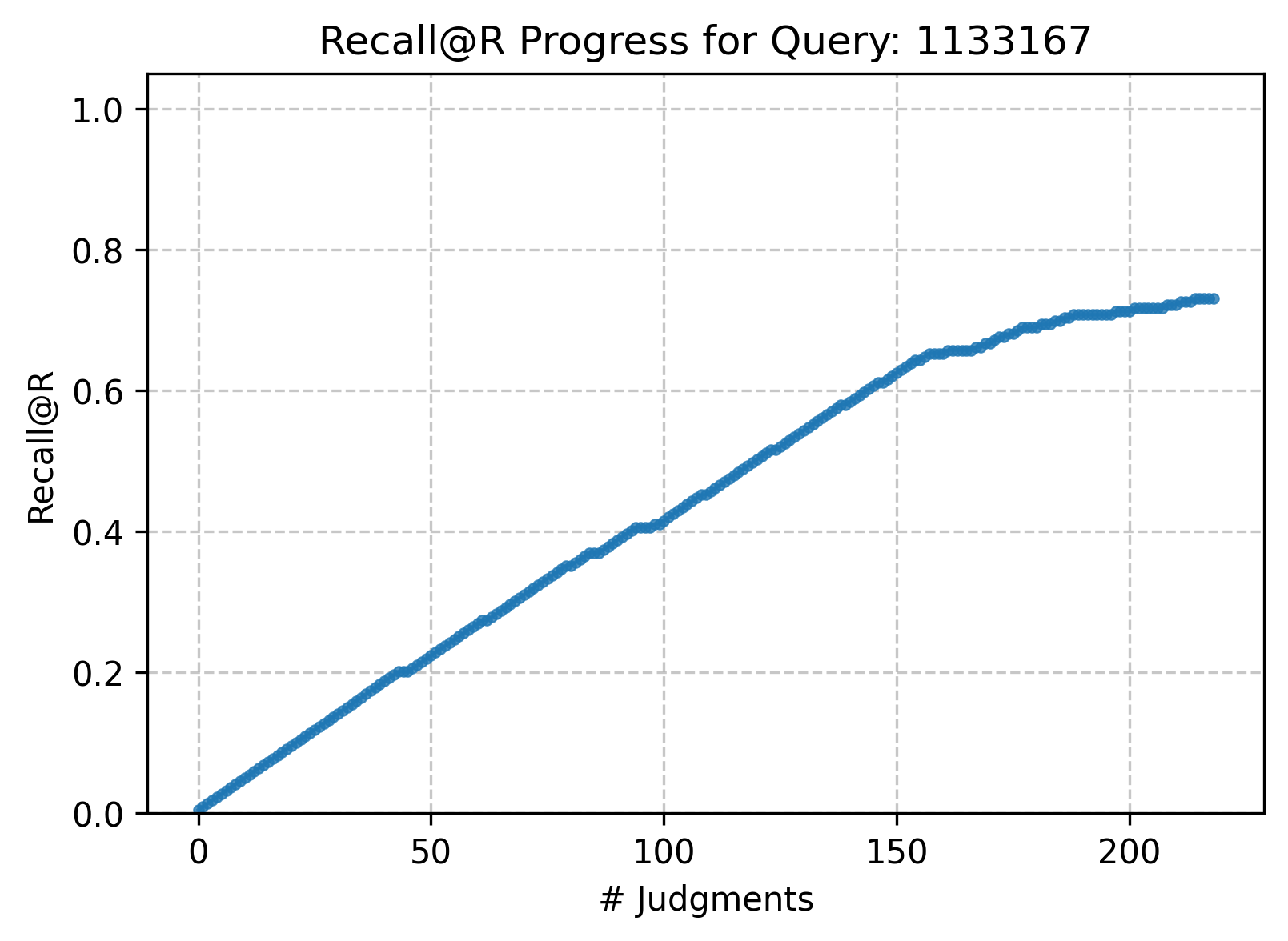} 
    \caption{An example number of judgments made versus recall@R for QID \texttt{1133167} under Splade and Query Reformulation then MTF policy.}
    \label{fig:recall_progress}
\end{figure}

We report recall@$R$ for all 16 runs in Figure \ref{fig:recall_comparison}.
The figure illustrates how policies can be compared.
For example, the best overall policy (SPLADE \& Query Reform $+$ MTF) substantially outperforms every policy using BM25 as its sole ranking method, with no overlap in the reported 95\% bootstrap confidence intervals.
More generally, the figure shows that both the underlying ranking method and the policy used to explore ranked lists can materially affect performance, supporting the usefulness of a common framework for comparing such strategies.
As a more specific example, Figure \ref{fig:recall_progress} shows how recall@$R$ changes with the number of judgments made under SPLADE and query reformulation followed by an MTF policy for a specific query.

\begin{figure}[t]
    \centering
    \includegraphics[width=0.5\textwidth]{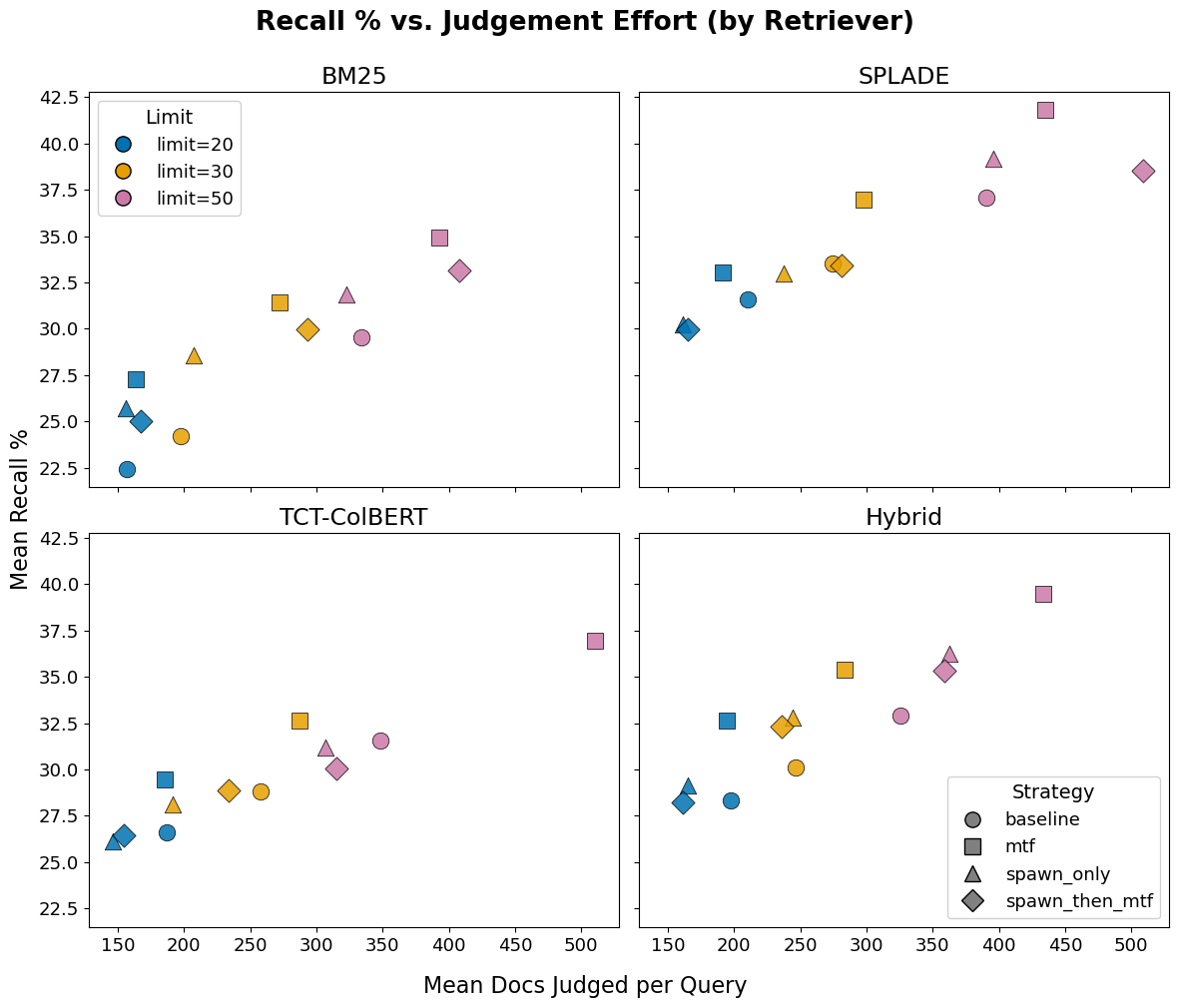} 
    \caption{Mean recall (\%) versus mean number of LLM relevance judgments consumed per query, for each retrieval pipeline (panels), judging budget (color), and policy (marker shape), evaluated under a global judging budget.}
    \label{fig:recall_vs_judgement_effort}
\end{figure}

Figure \ref{fig:recall_vs_judgement_effort} complements this comparison by isolating the cost of judging itself. Rather than scaling the budget by the number of known relevant documents $R$ for each query, here we fix a \textit{global} budget: a run stops once $\textit{limit} \in \{20, 30, 50\}$ consecutive newly-judged irrelevant documents have accumulated, independent of $R$. Because no official qrels are consulted at this granularity, recall is instead computed relative to the pooled union of documents that any policy, retriever, and budget judged relevant under the LLM judge, following standard pooling practice. This isolates how efficiently each policy converts LLM judgments into discovered relevant documents, holding the judge (GPT-4o-mini) fixed.

Query Reformulation then MTF achieves the highest recall for every retrieval pipeline and budget, but this gain is not free: because MTF interleaves judging across the original query and its reformulations, it can shift away from a list once it starts missing and draw a fresh hit from another list instead. Each such hit resets the single global consecutive-irrelevant counter, so the same global stopping criterion is triggered later than under Adhoc Search, which has only one list to exhaust. Forward Only Spawning and Bounded Spawning fall between Adhoc Search and Query Reformulation then MTF on both recall and judging cost, suggesting that on-the-fly divergent query generation improves recall over a single ranked list but is less judgment-efficient than committing to a batch of reformulations upfront. Adhoc Search is the least efficient policy under every retrieval pipeline, indicating that judging a single ranked list top-down makes poor use of a fixed judging budget. The retrieval pipeline matters as much as the policy: SPLADE and the hybrid pipeline recover more relevant documents per judgment than BM25 or TCT-ColBERT under every policy, mirroring the pattern already observed for recall@$R$ in Figure~\ref{fig:recall_comparison}.

\section{The Future: Learning to Search}

Learning to rank provides a useful analogy for our present purpose.
Treating ranking as a learning problem gave the field a more systematic way to specify objectives, compare methods, and make progress across a broad literature~\cite{joachims2002optimizing,burges2005learning,liu2009learning}.
While some prior work has also considered learning over longer search interactions and sessions~\cite{luo2015reinforce,luo2015direct,smucker2016switching}, LLM-guided search now seems to call for a similarly compact way of describing the search process itself.
As these systems become more capable, it becomes increasingly important to understand what counts as an improvement and why.
Our goal is to provide that framework, creating a principled basis for evaluation and comparison.

The policies described above illustrate how a variety of existing methods can be expressed within the spawn/select framework.
In this sense, the framework plays a role analogous to that of learning to rank: Before better methods can be learned, the search process itself must be described in a form that makes evaluation and systematic improvement possible.
The purpose of this paper is not to propose new learning algorithms, but to provide that evaluation formalization for LLM-guided search.

Within the framework of Section~\ref{sec:framework}, a search policy $\pi(\theta_t)$ determines when to execute $\textit{spawn}$ and which ranking to $\textit{select}$.
The reward formulation of Equation~\ref{eqn:reward} then provides a natural objective for optimizing this policy.
Rather than specifying heuristics such as MTF or fixed spawning rules, future work can instead learn policies that directly optimize the desired reward signal.
That reward may reflect the discovery of relevant documents under a judging budget, but it may also be defined in terms of information coverage, such as the discovery of informational nuggets or aspects of an information need
This paper uses the language of sequential decision making to define the problem clearly, but the work of learning strong search policies remains for future work.

Our framework makes it possible to ask a range of concrete learning questions about search policy.
For example, a learned policy may decide when it is worth spawning a new ranking, which existing ranking should be explored next, how effort should be allocated across competing search trajectories, and when the search has reached sufficient coverage to stop.
Different reward definitions then lead naturally to different search behaviors: one policy may emphasize finding as many relevant documents as possible within a judging budget, while another may emphasize coverage of informational nuggets or aspects of an information need.
In this sense, the framework does not itself solve learning to search, but it provides a representation in which such questions can be posed and evaluated systematically.

Viewed in this way, modern search agents performing retrieval-augmented generation or ``deep research'' can be interpreted as policies operating within this framework.
Queries are generated, documents are examined, information is accumulated, and the process continues until the system determines that the search has reached sufficient coverage.
The spawn/select abstraction therefore provides a foundation for learning search policies that control how information is gathered, allowing systems to optimize not only which documents to read, but how the search itself unfolds.
Rather than designing search strategies by hand, the system can learn how to explore a collection efficiently while prioritizing the documents and information that matter most.

\section{Conclusion}

We presented a sequence decision framework for LLM-guided search that puts these search strategies on a common footing, making them easier to describe, compare, evaluate, and eventually improve.

In our framework, a search process is modeled as a policy operating over a simple environment in which each step produces
an observation from a document.
The policy chooses between two actions: $\textit{select}$, which continues to explore an existing ranking, and $\textit{spawn}$, which generates a new ranking through query reformulation, reranking, or model-based retrieval strategies.
Expressing search in terms of these two actions provides a compact way to describe a wide range of existing approaches.

Using our framework, classic high-recall retrieval and test collection development strategies such as Move-to-Front pooling, Interactive Searching and Judging, and Continuous Active Learning can all be interpreted as policies within a common framework.
The same formulation also shows how ideas from bandits and reinforcement learning might eventually be applied to the design of search policies.
This perspective is particularly relevant for modern LLM-guided systems, in which large language models can generate queries, judge documents, and identify informational nuggets, enabling increasingly autonomous search loops.
More broadly, the framework suggests a shift in perspective: rather than designing search strategies by hand, we can begin to study how search policies themselves should be represented, compared, and ultimately learned.

\bibliographystyle{ACM-Reference-Format}
\bibliography{paper}

\newpage

\section*{Appendix} %
\appendix

\section{Batch Query Reformulation Prompt} \label{app:batch_query_reform}
\begin{lstlisting}[language=Python, breaklines=true, basicstyle=\small\ttfamily, frame=single]
class BatchReformSig(dspy.Signature):
    query = dspy.InputField(
        desc="Original user search query"
    )
    docs = dspy.InputField(
        desc="List of retrieved documents for context"
    )
    improved_queries = dspy.OutputField(
        desc="JSON-parsable list of 8 distinct search queries. These can be better phrasings of the original intent or targeted queries for different sub-aspects, depending on what will best improve retrieval."
    )
\end{lstlisting}

\section{Iterative Divergent Search Prompt} \label{app:divergent_sig}
\begin{lstlisting}[language=Python, breaklines=true, basicstyle=\small\ttfamily, frame=single]
class IterativeDivergentSearchSig(dspy.Signature):
    original_query = dspy.InputField(
        desc="Original user search query"
    )
    original_top_retrieved_docs = dspy.InputField(
        desc="Top docs for original query"
    )
    search_history = dspy.InputField(
        desc="Json list of past search history: each query searched and the corresponding top retrieved documents with their relevance score."
    )
    exploratory_query = dspy.OutputField(
        desc="A novel query designed to maximize retrieval of unique, previously unseen documents relevant to the original query."
    )
\end{lstlisting}

\end{document}